\documentstyle[12pt]{article}
\newcommand{\LI}{\hbox to\hsize}
\newcommand{\LLI}[1]{\LI{#1\hss}}

\newcommand{\PWm}{\mbox{$\rm W^{-}$}}

\newcommand{\PM}[1]%
{\mbox{$m_{\rm #1}$}} 
\newcommand{\BEQ}{\begin{equation}}
\newcommand{\EEQ}{\end{equation}}

\newcommand{\incircle}[1]{\mbox{{\hbox{$\bigcirc$}\kern-0.7em
\lower0.05ex\hbox{\mbox{{\scriptsize\rm #1}}}}}}
\newcommand{\eq}[1]{eq.~(\ref{#1})}
\newcommand{\VIZ}{\mbox{\em viz.\/ }}
\newcommand{\CF}{\mbox{\em cf.\/ }}
\newcommand{\IE}{\mbox{\em i.e. \/}}

\newcommand{\EG}{\mbox{\em e.g.\/ }}

\newcommand{\TIT}[1]{\vskip5mm \begin{center} {\Large\bf #1 }\\[10mm]
G. Domokos and S. Kovesi--Domokos
\\
The Henry A. Rowland Department of Physics and Astronomy \\
The Johns Hopkins University \\
Baltimore, MD 21218\footnote{E--MAIL: SKD@HAAR.PHA.JHU.EDU}
                   \end{center}\vskip8mm} 
\newcommand{\ABS}[1]{{\small #1}\vskip5mm} 
\begin{document}
\TIT{Field Theoretic Description of High Energy Neutrino Interactions in
Matter}
\ABS{In this paper we begin the development of a formalism for the
description of high energy neutrino interactions. It is based
upon  field theory quantized on a  null plane.
 We set up the general formalism as well as some techniques needed to
perform phenomenological calculations. We show that the formalism
developed by Wolfenstein is recovered at the cost of making two approximations:
one has to treat the charged lepton fields in the Hartree--Fock approximation
and one has to take the short distance limit of the Hartree--Fock correlation
function. As an example, we discuss the resonant interaction of
electron neutrinos in an electron gas.}
\section{Introduction}
The theory of neutrino interactions at high energies will play an
increasingly important role in the future. High energy neutrino
oscillation and interaction experiments will be performed by means of
 accelerator generated neutrino beams as well as in
experiments relying upon extraterrestrial sources of neutrinos such
as active galactic nuclei (AGN) and binary systems of stars.
In the latter setup, the center of mass energies in a neutrino
interaction with matter are likely to reach a few TeV, corresponding
to laboratory energies of the incident
neutrinos of the order of a few PeV.
Traditionally, neutrino interactions have been either described
by the single gauge boson exchange approximation (as in the case
of the theory of deep inelastic scattering of neutrinos on quarks)
or in a contact interaction approximation (for instance, in the theory
of neutrino oscillations in matter or in vacuum as developed
by Wolfenstein, \cite{wolfenstein}.) While there exist several
derivations of the Wolfenstein formalism~\cite{Bethe, Mannheim},
it is clear that a  more general formalism is needed  at the
highest energies. To quote but one example, if electron
antineutrinos of an energy of about 6.4 PeV pass through an electron
gas, they excite the \PWm resonance (the ``Glashow resonance'').
Such a situation may occur, for instance, when electron antineutrinos
generated deep in an AGN pass the electron plasma on their way out.
Clearly, neither the contact interaction approximation, nor a
gauge boson exchange approximation are adequate; the resonance
is excited in the s--channel; in this case, one has to be
able to treat the presence of matter adequately. Due to the fact
that neutrinos are nearly massless and we are interested in their
interactions at high energies, the null plane or front form formulation
of the theory is the most convenient one for the purpose. In the next
section we briefly review the formalism. Thereafter, we outline the procedure
for obtaining an effective theory of neutrino interactions in matter; as an
application, we show how  Wolfenstein's equation is recovered form the
general theory. In section 4 we illustrate the use of the theory
by describing the interaction of electron antineutrinos ($\overline{\nu_{e}}$)
with an electron gas in the energy region where the W boson is excited as
an s channel resonance. Sec. 5 contains a discussion of the results.

\section{A Review of the Null Plane Formalism}
The null plane or front form of quantum field theory is, in essence,
a constrained Hamiltonian formulation of a field theory given by
its Lagrangian and either operator quantization rules or
path integral prescription. Instead of a spacelike surface, however,
as it is the practice in setting up a traditional Hamiltonian formalism, one
prescribes Cauchy data on a plane with a null normal vector.
Such planes always contain characteristic lines of a relativistic wave
equation. As a consequence, the number of independent
Cauchy data is smaller than the ones one can prescribe  on a spacelike
surface, see, \EG~\cite{hadamard}. This is expressed in the form of
constraints obeyed by the fields entering the theory. As in any Hamiltonian
formulation of a relativistically invariant field theory, manifest Lorentz
invariance is lost. In what follows, we describe the formulation of
the theory of a  Dirac fermion interacting with an external gauge
field. The Dirac field may carry a representation of
an internal symmetry group. The generalization to the case
of a Yang--Mills theory or a
theory containing scalar fields as well is straightforward and it
has been described in a number of articles on the subject; a sampling of
some articles is given in ref.~\cite{nullplane}. Recently, an attempt has been
made to restore some of the symmetries (parity invariance in particular)
in a null plane formulation, see~\cite{jacob}. Due to the fact that
we are interested in a parity violating theory, those developments are not
needed here.

We begin with introducing a coordinate transformation in Minkowski
space. We always work with a metric chosen as $g_{\mu \nu}
= diag (1, -1, -1, -1)$ in the usual Cartesian basis.
Now introduce the coordinates,
\BEQ
t = \frac{1}{\sqrt{2}}\left( x^{0} - x^{3}\right) \hskip1.5mm
{\rm and}  \hskip1.5mm
z= \frac{1}{\sqrt{2}}\left( x^{0} + x^{3}\right).
\label{eq:nulldirections}
\EEQ
Correspondingly, the components of the metric tensor become,
\BEQ
g_{zt} = g_{tz} =1, \hskip 0.5mm g_{AB} = - \delta_{AB}
\label{eq:metric}
\EEQ
and all other components vanish. Here and in the following, capital
Latin subscripts and superscripts refer to the directions
perpendicular to the chosen conjugate null directions given by
\eq{eq:nulldirections}.
Correspondingly, the generators of the Dirac algebra are,
$ \gamma^{t}=  \left(\gamma^{0} + \gamma^{3}\right)/\sqrt{2},\hskip1em
\gamma^{z}= \left(\gamma^{0} - \gamma^{3}\right)/\sqrt{2}
 \hskip1.5mm {\rm and}  \hskip1.5mm \gamma^{A}$.
The Dirac Lagrangian is of the usual form,
\BEQ
{\cal L} = \overline{\psi}\left(\frac{i}{2}\partial
- ig_{(+)}A_{(+)} -ig_{(-)}A_{(-)}\right)\psi
+ \overline{\psi}m\psi
\label{eq:dirac}
\EEQ
Here, $\partial : =
1/2 \gamma^{\mu} \stackrel{\leftrightarrow}{\partial}_{\mu},
A_{\pm} := \gamma^{\mu}\left(1\pm \gamma_{5}\right) A^{(\pm)}_{\mu}$.
Internal symmetry indices are suppressed.

We now introduce spinor projectors corresponding to the two null
directions given in \eq{eq:nulldirections}, \VIZ
\BEQ
P_{t} = \gamma_{t}\gamma^{t}  \hskip1.5mm {\rm and}  \hskip1.5mm P_{z} =
\gamma_{z}\gamma^{z}.
\label{projectors}
\EEQ
For the sake of brevity, we also introduce the notation,
\[
P_{t}\psi = \phi  \hskip1.5mm {\rm and}  \hskip1.5mm P_{z}\psi = \chi .
\]
A straightforward manipulation leads to the useful relation:
\BEQ
\gamma_{5} =  -i\gamma^{1}\gamma^{2}\left( P_{t} - P_{z}\right).
\label{helicity}
\EEQ
Consequently, the chiral projectors, $\left(1 \pm \gamma_{5}\right)/2$
act as helicity projectors upon the spinors projected to
the conjugate null directions given by
\eq{eq:nulldirections}. We define  the helicity projectors,
\BEQ
H^{\pm} = \frac{1 \mp i \gamma^{1}\gamma^{2}}{2}
\EEQ
With this,  \eq{eq:dirac} becomes:
\begin{eqnarray}
\sqrt{2}{\cal L} & = & i \phi^{\dag}\partial_{t}\phi
+ i \chi^{\dag}\partial_{z}\chi + \phi^{\dag}\gamma^{z}m\chi
+ \chi^{\dag}\gamma^{t}m\phi \nonumber \\
 & - & i \phi^{\dag}\left( g_{(+)}A^{(+)}_{t}H^{+}
+ g_{(-)}A^{(-)}_{t}H^{-}\right)\phi \nonumber \\
 & - & i \chi^{\dag}\left( g_{(+)}A^{(+)}_{z}H^{-}
+ g_{(-)}A^{(-)}_{z}H^{+}\right)\chi \nonumber \\
  & +  & \phi^{\dag}\gamma^{z}\gamma^{A}\left(
\frac{i}{2}\partial_{A} - i g_{(+)} A^{(+)}_{A}H^{-}
- i g_{(-)}A^{(-)}_{A}H^{+}\right)\chi \nonumber \\
 & + & \chi^{\dag}\gamma^{t}\gamma^{A}\left( \frac{i}{2} \partial_{A}
-i g_{(+)} A^{+}_{A}H^{+}
- i g_{(-)} A^{-}_{A}H^{-}\right)\phi
\label{nulldirac}
\end{eqnarray}

It is to be emphasized that \eq{nulldirac} is completely symmetric
under the interchange of the two conjugate null directions.
However, the symmetry is destroyed if we decide to solve an initial value
problem by specifying the Cauchy data on one of the null planes.
For the sake of definiteness, we describe the procedure by specifying
initial data on a plane $t=0$; the procedure for Cauchyi data specified
on $z=0$ is completely analogous to the one described here and it can be
obtained by interchanging $\phi$ with $\chi$.

We recognize that if $t$ is regarded as the ``time'' variable,
only $\phi$ obeys an equation of motion; there is no time
derivative in the equation obeyed by $\chi$. Consequently,
the equation obeyed by $\chi$ is a constraint and thus $\chi$
can be eliminated altogether from the equations of motion.
Given the the Lagrangian, \eq{nulldirac}, the constraint can be
solved at least formally. The resulting equations of motion obeyed by $\phi$
are, in general, non--local; nevertheless,
they are legitimate dynamical equations.

In the case of \eq{nulldirac}, we proceed by choosing the gauge,
$ A^{\pm}_z = 0$. In this gauge the solution of the constraint is trivial,
since one just has to invert the operator $\partial_{z}$. In order to present
the result, we introduce the  covariant derivatives,
\begin{eqnarray}
\nabla^{(\pm)}_{k}&  = & \partial_{k} -  g_{(\pm)}A^{\pm}_{k}\nonumber \\
  &  & (k= A, t)
\label{covariant}
\end{eqnarray}
After elimination of the constraints, the end result reads:
\begin{eqnarray}
1/\sqrt{2} {\cal L} & = & \phi^{\dag} i\nabla^{(-)}_{t} H^{-}\phi
+ \phi^{\dag} i\nabla^{(+)}_{t} H^{+}\phi \nonumber \\
 & - & \left( \left( m - i \nabla^{(-)A}\right) H^{-}\phi \right)^{\dag}
\frac{{\cal P}}{k}\left( m - i \nabla^{(-)_{A}}\right) H^{-}\phi\nonumber \\
   &  -  &\left(  \left( m - i \nabla^{(+)A}\right) H^{+}\phi \right)^{\dag}
\frac{{\cal P}}{k}\left( m - i \nabla^{(+)_{A}}\right) H^{+}\phi
\label{nonrelativistic}
\end{eqnarray}

In this equation, we denoted $k= i\partial_{z}$ and, indeed,
{\em in the gauge chosen} the easiest
way of eliminating the  constraint is by means of a Fourier transformation
in $z$. The singularity at $k=0$ has to be eliminated by taking the
pricipal value, due to Hermiticity requirements.

We notice that \eq{nonrelativistic} is of the canonical form,

\BEQ
{\cal L} = \pi \partial_{t}\phi - {\cal H},
\label{jackiw}
\EEQ
as described recently by Jackiw~\cite{jackiw}. In fact, the entire
procedure of elimination of the constraints follows the pattern
described in that reference.

Were there no mass terms and no mixing between left and right--handed
components, there would be a one to one correspondence between
particles (antiparticles) and negative (positive) helicities,
respectively. Therefore a two component theory as described by
\eq{nonrelativistic} would be exact. However, at sufficiently high
energies one expects that the amplitudes of the ``wrong'' helicity
components are suppressed by factors of the $O(m_{\alpha}/E)$, where
$m_{\alpha}$ is some of the eigenvalues of the mass matrix appearing
in the previous equations, for instance, in \eq{nonrelativistic}.
Therefore, the two component theory is expected to be a good effective
theory. Parity conservation
(see ref.~\cite{jacob}) is not an issue since we are dealing with electroweak
interactions.  (We recall that under parity, $t\leftrightarrow z$,
$x^{A}\rightarrow - x^{A}$ and $\phi \leftrightarrow \chi$. Thus,
specifying Caucy data on a $t=const.$ ($z=const.$, resp.) surface,
parity is, by necessity, violated.)

The Feynman rules can be easily read off from \eq{nonrelativistic}.
One notices that they have the appearence of the rules for
a nonrelativistic Schr\"{o}dinger theory in 2+1 dimensions. This
is not an accident: it is due to the fact that the stability group
of a null direction in Minkowski space is $E(2)$, the Euclidean
group in two dimensions.
\section{Neutrino Interactions in Matter}
In this section we begin the development of the formalism needed to
describe neutrino interactions in matter. For the sake of simplicity, we
discuss explicitly the case of neutrino interactions in a
uniform  electron gas. As
we proceed, it will become obvious that the formalism can be generalized in
a straightforward manner to describe other physical situations of interest.

Our starting point is the effective action, see, \EG
ref.~\cite{ramond}, $\Gamma$, vieved as a
functional of classical neutrino and electron fields, denoted by $\psi$ and
$\Psi$, respectively. Both $\psi$ and $\Psi$ are regarded as
Dirac fields. Internal
symmetry indices are suppressed as before.

The effective action can be expanded in a functional power series (Volterra
series). We argue that in many cases of
physical interest, the Volterra series of $\Gamma$ can be broken off
after the first few terms. Clearly, the the second derivatives with respect
to $\psi$ and $\Psi$ give the free actions for these fields, with the
appropriate mass terms. Fourth and higher derivatives give the effective
interaction kernels. Both from a dimensional argument and from the explicit
calculation that follows, one realizes that terms proportional to
$(\overline{\Psi}\Psi)^{k}$ (with any arrangement of the
space--time arguments of the fields) are proportional to $n_{e}^{k}$, $n_{e}$
being the density of the electron gas. As a consequence, in most cases,
terms proportional to higher powers of the density can be dropped\footnote{
This is, in essence, the classical statistical argument for the dominance
of binary collisions in a gas of moderate density.}. On a similar basis
we omit derivatives higher than second in $\psi$, because we assume that
the neutrino beam considered is sufficiently dilute so that self interactions
of neutrinos can be neglected. (In addition, in the electroweak theory
terms containing higher powers of the electron density also contain
higher powers of the fine structure constant and, perhaps, of inverse
gauge boson masses.)

With this in mind, we now write the effective action:
\begin{eqnarray}
\Gamma [\psi, \Psi] & = & S_{0}[\psi] + S_{0}[\Psi]
+ \int \overline{\Psi (1)}\overline{\Psi (2)}
{\cal H}(1,2;3,4 )\Psi (3) \Psi (4) \nonumber \\
  & + & \int \overline{\psi (1)}\overline{\Psi (2)}
{\cal K}(1,2;3,4 )\Psi (3) \psi (4) +\ldots
\label{effaction}
\end{eqnarray}

In \eq{effaction}, the space--time points have been denoted
simply by Arabic numerals; the integrations extend over points
occurrring twice under the integral sign.
Clearly, \eq{effaction} defines an interacting system of electrons and
neutrinos. Assuming that ${\cal H}$ and ${\cal K}$ have been computed
in some approximation,  one can solve this classical field theory in order to
represent the neutrino electron interaction. The relevant Green functions can
be represented as  a sum of tree diagrams, with interaction vertices given by
the kernels ${\cal H}$ and ${\cal K}$. An easy way of generating these
diagrams is to introduce an auxiliary second quantization of the electron
and neutrino fields and then compute the tree diagrams in this theory.

As an application, let us show how the Wolfenstein formalism is recovered
in this framework.

The crucial point is to make a Hartree--Fock approximation in \eq{effaction}.
This means that one has to replace bilinear products of the form
$\overline{\Psi (1)}\Psi (2)$ in the terms proportional to ${\cal H}$
and ${\cal K}$ (and in the higher order terms)
in the effective action by their expectation values
in the electron gas. In this way, the effective action reduces to a quadratic
functional. On performing the above--mentioned truncation on the term
proportional to ${\cal H}$, one can absorb the resulting functional
into $S_{0}[\Psi]$ to a good approximation. In fact, it follows
from straightforward invariance arguments that the expectation value
( denote it by ${\bf H})$ must be of the form,
\BEQ
{\bf H} =\int  \overline{\Psi}h_{0}\Psi  + \overline{\Psi}h_{1}\partial \Psi
+ \partial_{\mu}\overline{\Psi}h_{2}\partial^{\mu}\Psi + \ldots \hskip0.3em,
\label{electrons}
\EEQ
where all fields are to be taken at the same point.

If the electron gas is non--relativistic, as we assume, the higher
derivative terms  may be omitted and we see that the effect of the
self interaction within the electron gas is just a mass shift and a wave
function renormalization. (In standard many body theory, see \EG
ref.~\cite{fetterwalecka}, one often introduces an energy dependent
{\em effective mass} in order to take some of the higher derivative terms
into account. While this can be done within the framework of
the present formalism, it is unimportant from the point of view of the
argument that follows.)

Let us now concentrate on the electron--neutrino interaction term.
On  replacing the bilinear electron operator by its expectation value,
the resulting expression is of the form:
\BEQ
\int \overline{\psi (1)} {\bf\rm K}(1,2) \psi (2),
\EEQ
where the kernel is given by
\BEQ
{\bf\rm K}(1,2)  =  \int {\cal K}(1,3;2,4)C(3,4).
\EEQ
The quantity denoted by $C(3,4)$ is the two point correlation function of the
electron gas. Due to the homogeneity of the electron gas, it depends only
on the difference of its two arguments.
On making a non--relativistic approximation to the kinematics for quantities
referring to the electron gas, one obtains the well known expression:
\BEQ
C({\bf x}, t) = n_{e} \frac{3 \exp (i \tau m_{e}/p_{F})}{2 \rho}
\int_{-1}^{1}\! du u \sin u\rho \thinspace \exp(i u^{2} \tau p_{F}/m_{e})
\label{correlation}
\EEQ
In \eq{correlation}, $p_{F}$ is the Fermi momentum;
 $\tau$ and $\rho$ are the time and radial distance
measured in units of the Fermi wavelength, \IE $\tau = t p_{F}, \hskip1em
\rho =
r p_{F} $. (In the usual nonrelativistic treatment, the exponential factor
before the integral is absent; this is merely a question of the
definition of the
chemical potential.) Finally, $m_{e}$ stands for the value of the effective
electron mass in the gas.

One can now proceed to decompose the Dirac spinor describing the neutrino
according to its projections onto the the two conjugate null
directions as described in the previous Section and eliminate the constraint.
In order to present the result, we use an operator notation, such that,
for instance:
\[
{\bf\rm K}\phi := \int {\bf\rm K}(1,2)\phi(2)
\]
In this way we obtain the effective Lagrangian:
\begin{eqnarray}
\sqrt{2} L_{eff} & = & i \phi^{\dag}\partial_{t}\phi \nonumber \\
  &  - & \phi^{\dag}\left( - i \nabla -m + K\right)
\left(i\partial_{z} +\gamma^{t}K\right)^{-1}
\left( i\nabla -m +K\right)\phi ,
\label{efflagrangian}
\end{eqnarray}
where $\nabla = \gamma^{A}\partial_{A}$.

In essence, the variation of \eq{efflagrangian} yields the Wolfenstein
equation describing the propagation of a neutrino in a medium.
In order to obtain the form usually quoted in the literature, \EG
in ref.~\cite{kuopantaleone}, the following steps are needed.
\begin{enumerate}
\item One assumes that the propagation is one dimensional; hence,
$\partial_{A}\phi$ can be chosen to be zero by an appropriate choice
of the coordinate system.
\item The interaction in the medium is given by the standard model
of electroweak interactions, \IE in the rest frame of the gas,
\BEQ
K = K_{v}\gamma^{0} + K_{a}\gamma^{0}\gamma^{5}
\EEQ
\item One neglects the ``wrong helicity'' components, \IE one
approximates,
\[  H^{-}\phi \approx \phi , \hskip1em H^{+}\phi \approx 0
\]
and ({\em mutatis mutandis}) similarly for the conjugate problem
in which $z$ is regarded the time  and $\chi$ the dynamical
variable.
\item One realizes that in \eq{correlation} the characteristic momentum
scale is given by $p_{F}$, while in $K$ it is the gauge boson mass, $M$.
In any environment of interest (with the possible exception of the
very early Universe) $M\gg p_{F}$.

Hence, one can safely approximate,
\[
C({\bf x},t)\approx C(0,0) = n_{e}
\]
\end{enumerate}

As a result, considerable simplifications occur.
In particular, after neglecting the ``wrong'' helicities and replacing the
the electron correlation function by the electron density, one realizes that
the coefficients $K_{v}$ and $K_{a}$ occur only in the combination:
\BEQ
 \kappa = n_{e}\left(K_{v} - K_{a}\right)
\label{kappa}
\EEQ
By eliminating the constraint, one identifies the the effective
Hamiltonian for the field $\phi$:
\BEQ
H_{eff}= \phi^{\dag}\left[ \frac{m^{2} + \nabla^{2}}{2 (k - \kappa)}
+ \kappa \right]\phi
\label{Heff}
\EEQ
If the denominator in \eq{Heff} were just equal to $k$, this would be
identical to the Hamiltonian for the Wolfenstein equation: assuming
one dimensional propagation and choosing the coordinate system appropriately,
one would obtain the equation discussed \EG in ref.~\cite{kuopantaleone}.
Under certain circumstances, one is indeed justifed in approximating
$k-\kappa \approx k$. However, the correct formula is given by
\eq{Heff} and care is needed in cases when the denominatior may become small.

The exercise just described is useful because it  makes the
limitations of the Wolfenstein formalism explicit. Within the framework
where
it is usually applied, \IE  the treatment of the solar neutrino
problem, the limitations are totally insignificant. However, often
one finds applications of the formalism to problems involving
high energy neutrino interactions, where it is, at best, of  limited
validity. In particular, one notices that the Hartree--Fock
approximation one has to make in order to reproduce Wolfenstein's
results describes the interaction of neutrinos with an
electron gas entirely in terms of the creation and subsequent annihilation
of a hole in the Fermi sea. At energies of interest in {\em high}
energy neutrino interactions, with $\sqrt{s}$ perhaps a few hundred
GeV, large momentum transfer processes are important; in particular,
electrons can be ejected into the continuum instead of annihilating with
a hole.
\section{Neutrino Interactions at the W Resonance}
As an illustration of the
formalism developed in the preceding Section,
we  consider $\overline{\nu_{e}}$ of a
laboratory energy approximately equal to 6.4 PeV,
incident upon an electron gas. In a previous paper,~\cite{resonant} we
discussed the physical circumstances under which this process leads to
somewhat unanticipated results. However, there we used the Wolfenstein
formalism uncritically. Here we show that the present formalism reproduces the
the results of ref.~\cite{resonant}.

We notice that at the energy mentioned, W is excited as an s--channel
resonance. As a consequence, it is reasonable to retain only those
diagrams which resonate at the W mass. For the sake of simplicity,
we also make the customary approximations, \VIZ we evaluate the self energy
part of the W propagator and the vertices at the the mass of the W.
(For most purposes, this is an adequate approximation and, in essence, it is
equivalent to using a Breit--Wigner formula to describe the resonance.)

In this approximation the effective electron neutrino interaction
can be written in the
form:

\begin{equation}
\Gamma_{int} = g^{2} \int \overline{\Psi(1)}\gamma_{\mu}
\frac{1-\gamma_{5}}{2}\psi(1)\Delta \left(1 - 2\right)
\overline{\psi (2)}\gamma^{\mu}\frac{1-\gamma_{5}}{2}\Psi(2)
+ (conj.),
\end{equation}
where $g^{2}$ is the coupling evaluated at the resonance
 and $\Delta$ stands for the propagator of the W: it is
the Fourier transform of the quantity,
\[ \frac{1}{M^{2} - k^{2} - iM\gamma} .\]
As usual, we denoted the self energy part evaluated at resonance
by $M^{2} - i M\gamma$. Hence, $M$ is the physical mass and
$\gamma$ is the total width.
By performing a Fierz transformation, the last equation
can be brought to the more convenient form:
\BEQ
\Gamma_{int} = g^{2} \int \overline{\Psi(1)}\gamma_{\mu}
\frac{1-\gamma_{5}}{2}\Psi(2)\Delta \left(1 - 2\right)
\overline{\psi (2)}\gamma^{\mu}\frac{1-\gamma_{5}}{2}\psi(1)
+ (conj.)
\EEQ
In this form one can easily make a Hartree--Fock approximation, by
replacing the bilinear quantity in the electron operators by
its expectation value. On assuming that the
electron correlation function is evaluated in the
rest frame of the gas and replacing the correlation
function by its value at the origin, the contribution is of the form as in
\eq{kappa}, with $K_{v}=-K_{a}$.
At this pont, we are now ready to write down the effective Hamiltonian with
the resonant interaction. Due to the fact that the effective action was reduced
to a quadratic functional and we consider one dimensional propagation only,
it is convenient to Fourier transform the effective action. In this manner,
the computation of the inverses of the operators entering the constraints
becomes trivial. We  write the Hamiltonian
in the form (\CF \eq{Heff}):
\[
H = \phi^{\dag}\frac{m^{2} + 2(p - \kappa)\kappa}{2(p - \kappa)}\phi.
\]
Here, $p$ stands for the variable conjugate to $z$; we suppressed
the argument of $\phi$; integration over $p$ is understood.

The question arises whether this Hamiltonian can be replaced by a
conventional one, \IE
\[
H = \phi^{\dag}\frac{m^{2} + 2p \kappa}{2p}\phi.
\]
We now proceed to show that in the case of the resonant interaction
discussed here, such a replacement is well justified. First of all,
it is convenient to rewrite the expression of $\kappa$ in terms
of the elastic and total widths of the resonance. This is easily
accomplished by noticing that the usual invariant variable $s$
can be written as $s= 2p{\rm m_{e}}$, ${\rm M_{e}}$ being the mass of
the electron. Straightforward manipulations then lead to the
expression near $s= M^{2}$:
\BEQ
2p \kappa \approx  m_{c}^{2} \frac{M\gamma_{e}}{s - M^{2} + i M\gamma}
\label{potential}
\EEQ
In \eq{potential}, $m_{c}$ is a mass scale characterizing the electron
gas; it is defined by the relation $m_{c}^{2} = n_{e}/{\rm m_{e}}$.
The following Table, taken from ref.~\cite{resonant},
contains the values of the characteristic mass for some
environments of interest.
\vskip4mm
\begin{minipage}{\hsize}
\begin{center}
{\bf Electron densities and characteristic masses \\ for some
environments}\\[3mm]
\begin{tabular}{|c|c|c|}\hline
Environment & $n_{\rm e} [{\rm cm}^{-3}]$ &${\rm m}_{c}^{2}
[{\rm eV}^{2}]$\\ \hline
stellar interior (sun) & $10^{27}$ & $2\times 10^{7}$\\
Earth & $1.6\times 10^{24}$ & $ 3\times 10^{4} $\\
water & $3\times 10^{23}$ & $5\times 10^{3}$\\ \hline
\end{tabular}
\end{center}
\end{minipage}
\vskip4mm
We now notice that there are no vanishing denominators in the Hamiltonian,
 since $\kappa$ is complex. Furthermore, the correction to
the conventional Hamiltonian arising from \eq{Heff} is of the order of
magnitude $2\kappa/p$. Near resonance this is
\[
\left| \frac{2\kappa}{p} \right| \approx \frac{\gamma_{e}m_{c}^{2}
{\rm m_{e}}^{2}}{4 \gamma M^{2}}.
\]
Even for an environment like a stellar interior, the corrections are
minuscule. There may be substantial corrections to the naive form of the
Wolfenstein Hamiltonian in very dense environments, \EG in the interior
of a neutron star or in the early Universe. However, in all probability,
the Hartree--Fock approximation breaks down before these corrections
become truly significant.

One can include flavor degrees of freedom without any difficulty and discuss
neutrino mixing near  the resonance as it has been done in ref~\cite{resonant}.
None of the conclusions about the validity of the Wolfenstein formalism is
altered by such a generalization and therefore we shall not dwell on this
topic any further.

\section{Discussion}

The basic purpose of  this work has been to establish a formalism leading
to an effectice theory of neutrino interactions in matter. It became
clear that the use of the null plane formalism is suitable for the
description of these interactions.
Clearly, however, what is
needed is a reliable calculation of the effective action. There are several
possible approaches to  this problem with an increasing degree of complexity.
In the approach discussed here, one puts the classical fields
corresponding to all but a few species equal to zero. (Here
we discussed the simplest case: only the fields corresponding
to neutrinos and the target medium are kept.) Even at this level,
some interesting generalizations are possible; in particular, at high
neutrino densities, such as in the early Universe, one can no longer
neglect the interaction between the various flavors of neutrinos;
consequently, one has to deal with a rather non--trivial
multichannel problem. In some recent articles,
Kostelecky, Pantaleone and Samuel,~ref.~\cite{pantaleonesam}
addressed this problem entirely within the framework of the
Wolfenstein formalism. However, those
authors, in essence, make  a
dilute gas approximation by neglecting
the effects of the Fermi sea, while keeping the interaction terms
between  neutrinos. The internal consistency of such a
procedure is not quite obvious and further studies are needed in
order to clarify the issues involved.

More importantly, in discussing high energy neutrino interactions,
one has to go beyond the Hartree--Fock approximation as already discussed,
since in that approximation a substantial amount of physical
information is lost. Furthermore, in order to obtain a reliable theory,
one will have to follow the evolution of several components of the
system even if ultimately only the transition probabilities for
the stable particles are kept.
\vskip4mm
\LLI{\bf Acknowledgement}
We benefitted from fruitful discussions with R.~Casalbuoni,
R.~Fletcher and L.~Madansky.

\end{document}